\documentclass[12pt,english]{article}
\usepackage[T1]{fontenc}
\usepackage[latin9]{inputenc}
\usepackage{float}
\usepackage{amsmath}
\usepackage{amsthm}
\usepackage{amssymb}
\usepackage{graphicx}
\usepackage{setspace}
\makeatletter

\newcommand{\noun}[1]{\textsc{#1}}
\providecommand{\tabularnewline}{\\}

\newcommand{\lyxaddress}[1]{
	\par {\raggedright #1
	\vspace{1.4em}
	\noindent\par}
}
\theoremstyle{plain}
\newtheorem{thm}{\protect\theoremname}
\theoremstyle{definition}
\newtheorem{defn}[thm]{\protect\definitionname}

\makeatother

\usepackage{babel}
\providecommand{\definitionname}{Definition}
\providecommand{\theoremname}{Theorem}

\begin{document}
\title{Quantum Neural Networks - Computational Field Theory and Dynamics}
\author{Carlos Pedro Gonçalves}
\maketitle

\lyxaddress{Lusophone University of Humanities and Technologies, p6186@ulusofona.pt}
\begin{abstract}
To address Quantum Artificial Neural Networks as quantum dynamical
computing systems, a formalization of quantum artificial neural networks
as dynamical systems is developed, expanding the concept of unitary
map to the neural computation setting and introducing a quantum computing
field theory on the network. The formalism is illustrated in a simulation
of a quantum recurrent neural network and the resulting field dynamics
is researched upon, showing emergent neural waves with excitation
and relaxation cycles at the level of the quantum neural activity
field, as well as edge of chaos signatures, with the local neurons
operating as far-from-equilibrium open quantum systems, exhibiting
entropy fluctuations with complex dynamics including complex quasiperiodic
patterns and power law signatures. The implications for quantum computer
science, quantum complexity research, quantum technologies and neuroscience
are also addressed.
\end{abstract}
\begin{description}
\item [{Keywords:}] Quantum Artificial Neural Networks; quantum neural
maps; quantum computing field theory; complex quantum systems.
\end{description}

\section{Introduction}

The connectionist paradigm for Artificial Intelligence (A.I.) played
a key role in the development of cybernetics and the complexity sciences
(Gorodkin \emph{et al.}, 1993; Dupuy, 2000; Novikov, 2016; Ivancevic
\emph{et al.}, 2018). The dynamics of networked computational systems
led, within the cybernetics paradigmatic basis, to the development
of an interdisciplinary link between dynamical systems science, computer
science, and evolutionary biology (Packard, 1988; Langton, 1990; Kauffman,
1990, 1991, 1993; Dupuy, 2000; Novikov, 2016; Ivancevic \emph{et al.}\emph{\noun{,
2018}}).

The next generation of cybernetics is quantum cybernetics which extends
the connectionist paradigm to the quantum framework, with quantum
artificial neural networks (QuANNs) as a main computational model
(Gonçalves, 2015a,b, 2017, 2019a,b, 2020; Novikov, 2016; Kwak \emph{et
al.}, 2021) which has recently been incorporated within the wider
context of quantum machine learning (Gonçalves, 2017; Beer \emph{et
al.}, 2020; Houssein \emph{et al.}, 2022; Parisi \emph{et al.}, 2022).

From a quantum computer science standpoint, a QuANN with \emph{n}
neurons and a two-level firing pattern can be addressed as an \emph{n}-qubits
quantum computing network, where the quantum computing gates are conditional
unitary operators that obey the network's connections, that is, the
unitary quantum computing operation associated with a given neuron
is conditional upon the input neurons' firing patterns, this leads
to an extension of the circuit model of quantum computation applied
to the quantum connectionist paradigm. Given the conditional gate
structure associated with each neuron, the final quantum circuit depends
upon the neuron activation sequence.

As shown in Gonçalves (2017), while a QuANN is capable of running
quantum algorithms and also of selecting algorithms depending on the
task, it can also operate as a quantum networked dynamical system,
with the computation performed having a dynamical signature at the
level of the quantum averages.

Major models of networked computation, such as cellular automata,
artificial neural networks (ANNs) and random Boolean networks (Packard,
1988; Langton, 1990; Kauffman, 1990, 1991, 1993; Wolfram, 2002), when
analyzed as dynamical systems led to the discovery of four major classes
of behavior:
\begin{itemize}
\item Class 1: steady state or fixed point dynamics; 
\item Class 2: periodic dynamics;
\item Class 3: random-like dynamics;
\item Class 4: an intermediate dynamics between classes 2 and 3. 
\end{itemize}
Class 4, also known as the \emph{edge of chaos}, was a key focus of
complexity research since it was shown that at the \emph{edge of chaos}
an evolutionary computing system maximized its fitness (Packard, 1988;
Langton, 1990; Kauffman, 1990, 1991, 1993), also, the \emph{edge of
chaos} dynamics seem to play a role in the conditions for the emergence
of complex noise resilient dynamics, since, at the \emph{edge of chaos},
a system is able to conserve an emergent order and at the same time
is capable of sustaining the necessary adaptive change (Kauffman,
1993).

In Gonçalves (2017), it was shown that QuANNs interacting with an
environment can lead to complex dynamics at the level of the mean
neural firing energy with \emph{edge of chaos}-like signatures.

To research on QuANNs as quantum dynamical systems, however, we need
to expand the theory to include a quantum field theory on the network,
this is the main objective of the present work, which is focused on
two major contributions:
\begin{enumerate}
\item The introduction of a complete formalism for QuANNs as quantum dynamical
systems (section 2) with:
\begin{itemize}
\item The definition of the networked computing formalism (subsection 2.1);
\item The formalization of the concept of a unitary quantum neural map,
which expands the concept of unitary quantum map, originally worked
within the context of quantum chaos theory (Stöckmann, 2000; Braun,
2001; Gonçalves, 2020) (subsection 2.1);
\item The expansion of quantum field theory to QuANNs, defining the concept
of a quantum neural field operator, and addressing the field's dynamics
as a function of the network's dynamics, with the formalization of
recurrence analysis methods to study the quantum field dynamics on
the neural network, we also exemplify the main theory with a specific
field, which is the neural activity field (subsection 2.2);
\item The application of von Neumann entropy to address the dynamics of
each neuron as an open quantum system in connection with the network
(subsection 2.3).
\end{itemize}
\item An implementation of section 2's theoretical framework to a two-neuron
quantum recurrent neural network (QRNN) and the study of the mean
neural activity field's dynamics and entropy dynamics (section 3).
\end{enumerate}
As we show in section 3, for the most elementary QRNN comprised of
two neurons, the quantum neural activity field's dynamics, for different
parameter values, can exhibit standard class 1 and class 2 dynamics,
for other parameter values brainwave-like patterns emerge for the
mean neural activity field dynamics with strongly correlated local
mean field values calculated as the quantum averages for the field
at each neuron, for other parameter values, the emergent dynamics
is class 4.

While the whole network undergoes a unitary evolution, there are entropy
fluctuations at the local neuron level, therefore, in conjunction
with the analysis of the dynamics for the mean neural activity field,
we study the resulting entropy sequences for each neuron, thus, treating
each neuron as an open quantum system.

For different parameter values, we find that the wave-like periodic
emergent pattern translates to an also emergent periodic pattern driving
the local (neuron-level) entropy dynamics, while for class 4 mean
neural activity field dynamics we also find class 4 dynamics in the
entropy dynamics, including power law signatures at the local neuron-level
entropy sequences with fluctuations never leading to maximum entropy
associated with a depolarized mixed density and fluctuations that
range from near zero entropy values to high but not maximum entropy,
therefore, each neuron operates as a far-from-equilibrium open quantum
system with no stabilized fixed decoherence pattern, a result that
may be key for the development of advanced networked quantum technologies,
since the local computing units do not tend to a maximum entropy and
recurrently return to near zero entropy values.

The implications of the work for quantum computer science, complexity
research, quantum technologies and neuroscience are addressed in section
4.

\section{Quantum Neural Networks, Computational Field Theory and Dynamics}

\subsection{Computational Structure of Quantum Artificial Neural Networks and
Quantum Neural Maps}

In order to produce a general computational framework on which to
discuss QuANNs, we need to consider a directed graph (digraph) structure
for the $n$ neuron network extended with a Hilbert space structure
and a set of conditional unitary operators, one operator for each
neuron, with the unitary computation conditional on the input neural
connections, formally this digraph can be defined as follows:
\begin{defn}
(QuANN) A QuANN is a digraph with the structure: 
\begin{equation}
G=\left(Q,D,\mathbb{H}_{Net},\Theta\left(\mathbb{H}_{Net}\right)\right),
\end{equation}
where $Q=\left\{ n_{0},n_{1},...,n_{n-1}\right\} $ is the set of
neurons, $D$ is the set of ordered pairs corresponding to the directed
edges, $\mathbb{H}_{Net}$ is the network's Hilbert space and $\Theta\left(\mathbb{H}_{Net}\right)=\left\{ U_{0},U_{1},...,U_{n-1}\right\} $
is a set of conditional unitary operators on $\mathbb{H}_{Net}$,
one for each neuron.
\end{defn}

The computational dynamics of QuANNs can be addressed as a system
of spinors on a network (Gonçalves, 2017). In the spinor model, when
each neuron can have just two base computational patterns of activity,
firing or nonfiring, corresponding to two energy levels, the Hilbert
space for the full network is given by $n$ tensor product copies
of the Hilbert space $\mathbb{H}_{2}$, which is spanned by the standard
qubit basis $B_{2}=\left\{ \left|0\right\rangle ,\left|1\right\rangle \right\} $,
where the ket vector $\left|0\right\rangle $ represents a nonfiring
neural activity and $\left|1\right\rangle $ a firing neural activity,
we are using the standard Dirac's notation where the ``ket'' vector
$\left|w\right\rangle $ represents a column vector and the ``bra''
vector $\left\langle w\right|$ is its conjugate transpose.

The Hilbert space for the network, in the case of a binary firing
pattern is defined as $\mathbb{H}_{Net}=\mathbb{H}_{2}^{\otimes n}$
and spanned by the firing pattern basis (Gonçalves, 2017):

\begin{equation}
B_{2}^{\otimes n}=\left\{ \left|s_{0},s_{1},...,s_{n-1}\right\rangle :s_{k}=0,1;k=0,1,...,n-1\right\} 
\end{equation}

A generalization of this model, for a finite number of logical states
encoded in quantum a neural firing activity, is obtained by expanding
the single neuron basis to $B_{l}=\left\{ \left|0\right\rangle ,\left|1\right\rangle ,...,\left|l-1\right\rangle \right\} $,
spanning the single neuron $l$-dimensional Hilbert space $\mathbb{H}_{l}$,
which would lead to the neural network's Hilbert space $\mathbb{H}_{Net}=\mathbb{H}_{l}^{\otimes n}$,
spanned by the generalized firing pattern basis:
\begin{equation}
B_{l}^{\otimes n}=\left\{ \left|s_{0},s_{1},...,s_{n-1}\right\rangle :s_{k}=0,1,...,l;k=0,1,...,n-1\right\} ,
\end{equation}
when $l=2$, this last basis reduces to the standard two-level firing
pattern basis. In this section, we work with the generalized formalism,
since it contains the two-level as a special case. In section 3, the
example is worked for $l=2$.

Now, the classical information states encoded in a neural firing dynamics,
formalized as basis vectors in $B_{l}^{\otimes n}$, are not the only
possible information states, we can also have a superposition of different
neural firing patterns, which can be formalized as a normalized ket
vector $\left|\psi\right\rangle $ on the network's Hilbert space,
expanded in the firing pattern basis as follows:

\begin{equation}
\left|\psi\right\rangle =\sum_{s_{0},s_{1},...,s_{n-1}}\psi(s_{0},s_{1},...,s_{n-1})\left|s_{0},s_{1},...,s_{n-1}\right\rangle 
\end{equation}

The squared modulus of each quantum amplitude $\psi(s_{0},s_{1},...,s_{n-1})$
provides for a statistical weight associated with each neural firing
pattern, so that, for an ensemble of identical independent neural
networks, the probability of the network exhibiting a specific firing
pattern $s_{0},s_{1},...,s_{n-1}$ is given by $|\psi(s_{0},s_{1},...,s_{n-1})|^{2}$. 

Different interpretations of quantum mechanics interpret this statistical
measure differently, for instance, in an Everettian interpretation
(Everett, 1957; 1973), each alternative firing pattern with a non-zero
amplitude corresponds, in this case, to a projected dimension of systemic
activity, with a projective intensity coinciding with the squared
norm of the projection which, in turn, corresponds to the squared
modulus of the amplitude $|\psi(s_{0},s_{1},...,s_{n-1})|^{2}$, so
that, considering an ensemble of identical independent neural networks
all described by the same multidimensional projective pattern (the
same ket vector), the ensemble density operator description leads
to a statistical measure (as a relative frequency) exactly coincident
with $|\psi(s_{0},s_{1},...,s_{n-1})|^{2}$, which means that, choosing
at random one network in the ensemble, the probability associated
with that network's exhibiting a given projected firing pattern coincides
with the squared modulus of the quantum amplitudes, this point was
addressed in Gonçalves (2015b).

Other interpretations of the probability link are possible (Cramer,
2016; Gonçalves, 2019a,b), we will not, however, assume here a specific
interpretation quantum mechanics, with the formalism holding for different
interpretations. The relevant point is the link to the probabilistic
description since further on we will need to work with quantum averages
for quantum fields on the network that rely on the above correspondence
between the squared modulus of the amplitudes and statistical weights
in an ensemble of identical independent networks.

Now, quantum computations on the network are formalized by way of
the operations of the unitary gates in $\Theta\left(\mathbb{H}_{Net}\right)$
on the normalized ket vectors $\left|\psi\right\rangle $, considering
this point, in order to introduce the concept of a unitary neural
map, let $p:\left\{ 0,1,...,n-1\right\} \mapsto\left\{ 0,1,...,n-1\right\} $
represent a permutation of the neuron indices, then, we can formalize
the concept of a unitary neural map as follows:
\begin{defn}
(Unitary Neural Map) A unitary neural map $F$ is defined such that,
given a permutation $p$ of neuron indices, the map is given by the
product:
\begin{equation}
F=U_{p(n-1)}...U_{p(1)}U_{p(0)}
\end{equation}
with $U_{p(k)}\in\Theta\left(\mathbb{H}_{Net}\right),k=0,1,...,n-1$.
\end{defn}

In this way, a unitary neural map is a product of the unitary gates
in $\Theta\left(\mathbb{H}_{Net}\right)$ in an order corresponding
to a neuron activation sequence that matches the permutation $p$.
Given a unitary neural map, we can define the sequence of iterations
of the map as:
\begin{equation}
\left|\psi(t)\right\rangle =F\left|\psi(t-1)\right\rangle ,
\end{equation}
which expands the unitary maps, worked in quantum chaos theory (Stöckmann,
2000; Braun, 2001), to the quantum neural computational setting.

In this case, $t$ represents the iteration step, and the QuANN's
dynamics is addressed in terms of a unitary quantum map which matches
the quantum computing circuit described by equation (5). Given equation
(6), and letting $\left|\psi(0)\right\rangle $ be the input for the
network, we get the output at iteration \emph{t} as:
\begin{equation}
\left|\psi(t)\right\rangle =F^{t}\left|\psi(0)\right\rangle 
\end{equation}

The above iteration scheme allows us to deal with QuANNs as quantum
networked dynamical systems. Now, in order to better address the network's
dynamics we need to introduce a quantum neural computing field theory.

\subsection{Quantum Neural Computing Field Theory}

QuANNs, when addressed as quantum networked dynamical systems, lead
to a bridge between quantum field theory and quantum computing, indeed,
to address QuANNs as quantum networked dynamical systems implies the
need to develop a quantum field theory on the network. Working with
the formalism introduced in the previous subsection, we can develop
a formalism for quantum fields on the network by introducing a general
quantum neural field operator $\alpha(k)$, as a field operator on
the network, formally:
\begin{defn}
(Quantum Neural Field Operator) A quantum neural field operator on
an \emph{n} neurons QuANN with $l$ firing levels $G=\left(Q,D,\mathbb{H}_{l}^{\otimes n},\Theta\left(\mathbb{H}_{l}^{\otimes n}\right)\right)$
is a field operator $\alpha(k)$ on the network defined with the following
structure:
\begin{equation}
\alpha(k)=\sum_{s=0}^{l-1}\alpha_{s}1_{2}^{\otimes(k-1)}\otimes\left|\alpha_{s}\right\rangle \left\langle \alpha_{s}\right|\otimes1_{2}^{\otimes(n-k)}
\end{equation}
where the coefficients $\alpha_{s}$ are real-valued and the projectors
$\left|\alpha_{s}\right\rangle \left\langle \alpha_{s}\right|$ project
over a basis $\left\{ \left|\alpha_{0}\right\rangle ,\left|\alpha_{1}\right\rangle ,...,\left|\alpha_{l-1}\right\rangle \right\} $
spanning the single neuron Hilbert space $\mathbb{H}_{l}$.
\end{defn}

Given the above definition, it follows that the operators commute,
and, for any normalized ket vector on the network of the form:
\begin{equation}
\left|\psi\right\rangle =\left|\phi\right\rangle \otimes\left|\alpha_{s}\right\rangle \otimes\left|\varphi\right\rangle 
\end{equation}
where $\left|\phi\right\rangle \in\mathbb{H}_{2}^{\otimes(k-1)}$
and $\left|\varphi\right\rangle \in\mathbb{H}_{2}^{\otimes(n-k)}$,
the following eigenvalue equation holds:
\begin{equation}
\alpha(k)\left|\psi\right\rangle =\alpha_{s}\left|\phi\right\rangle \otimes\left|\alpha_{s}\right\rangle \otimes\left|\varphi\right\rangle 
\end{equation}
for $s=0,1,...,l$.

Now, under the action of the quantum neural map, the field dynamics
can be adddressed in the Heisenberg picture as follows:

\begin{equation}
\alpha(k,t)=F^{\dagger}\alpha(k,t-1)F
\end{equation}
Therefore, in the Heisenberg picture, the field operators undergo
the unitary evolution while the vectors stay at their initial configuration,
so the unitary map's iteration rule applies to the field operator
rather than to the vector.

Assuming $\alpha(k,0)=\alpha(k)$, recursive application of the map
leads to the following link:
\begin{equation}
\alpha(k,t)=(F^{\dagger})^{t}\alpha(k)(F)^{t}
\end{equation}

Now, given the initial ket vector $\left|\psi(0)\right\rangle $,
the quantum average of the field at iteration step $t$ and at neuron
$k$, in the Heisenberg picture, is given by:
\begin{equation}
\bar{\alpha}(k,t)=\left\langle \psi(0)\left|\alpha(k,t)\right|\psi(0)\right\rangle 
\end{equation}
applying equation (12) we can transition from the Heisenberg to the
Schrödinger picture since:
\begin{equation}
\bar{\alpha}(k,t)=\left\langle \psi(0)\left|\alpha(k,t)\right|\psi(0)\right\rangle =\left\langle \psi(0)\left|(F^{\dagger})^{t}\alpha(k)(F)^{t}\right|\psi(0)\right\rangle ,
\end{equation}
which leads to the equivalent result for the quantum mean field value
at neuron $k$ obtained from the quantum averages calculated in the
Schrödinger picture:
\begin{equation}
\bar{\alpha}(k,t)=\left\langle \alpha(k)\right\rangle _{t}=\left\langle \psi(t)\left|\alpha(k)\right|\psi(t)\right\rangle ,
\end{equation}
so that both pictures lead to equivalent results.

Taking the sequence of real-valued quantum averages $\left\langle \alpha(k)\right\rangle _{t}$,
for an $n$ neuron network we can embed the sequence in $\mathbb{R}^{n}$,
so that we get a sequence of points $\left\langle \alpha\right\rangle _{t}=\left(\left\langle \alpha(0)\right\rangle _{t},\left\langle \alpha(1)\right\rangle _{t},...,\left\langle \alpha(n-1)\right\rangle _{t}\right)$,
which leads to a trajectory in $\mathbb{R}^{n}$.

Taking advantage of the Euclidean space metric topology of $\mathbb{R}^{n}$,
for any sample path $\left\{ \left\langle \alpha\right\rangle _{t}:t=t_{0},t_{0}+1,...,t_{0}+T-1\right\} $,
we can calculate the distance matrix $\mathbf{S}$ that stores the
distances for each pair of points, with entries defined as:
\begin{equation}
\mathbf{S}_{t,t'}=\left\Vert \left\langle \alpha\right\rangle _{t}-\left\langle \alpha\right\rangle _{t'}\right\Vert ,
\end{equation}
where $\left\Vert .\right\Vert $ stands for the Euclidean metric
defined on $\mathbb{R}^{n}$.

The distance matrix is symmetric of rank $T$, with the main diagonal
entries all equal to zero and it contains the information about recurrences
in a sample trajectory.

Taking advantage of the Euclidean metric topology of $\mathbb{R}^{n}$,
the pattern of recurrences can be extracted from the distance matrix
using a closed $\delta$-neighborhood structure, which leads to the
binary recurrence matrix $\mathbf{R}_{\delta}$ with entries:
\begin{equation}
\mathbf{R}_{t,t'}^{\delta}=\begin{cases}
0, & \left\Vert \left\langle \alpha\right\rangle _{t}-\left\langle \alpha\right\rangle _{t'}\right\Vert >\delta\\
1, & \left\Vert \left\langle \alpha\right\rangle _{t}-\left\langle \alpha\right\rangle _{t'}\right\Vert \leq\delta
\end{cases}
\end{equation}

The matrix $\mathbf{R}_{\delta}$ is, thus, binary and symmetric with
an entry containing the value 1 when two points in the sample trajectory
are not apart from each other more than $\delta$, in a closed Euclidean
neighborhood, which is the definition of a recurrence event, of course,
given this property the diagonal of the matrix, which corresponds
to the cases where $t=t'$, is comprised only of 1s.

We are using a closed neighborhood because it allows us to identify
fully periodic dynamics since, when the dynamics is fully periodic,
if the radius is set equal to zero, all diagonal lines parallel to
the main diagonal corresponding to the period in question will have
a value of 1 in each matrix entry, otherwise the value will be 0.
When the dynamics is not periodic there is a cutoff radius below which
we do not get any recurrences. In the nonperiodic case, diagonal lines
with a value of 1 in each entry, at a given radius, correspond to
a periodic or quasiperiodic skeleton that the dynamics revisits, these
are 100\% recurrence lines, that is, lines where the percentage of
points (recurrence matrix entries) that are recurrence points is equal
to 100\%. These lines are particularly important in identifying periodic,
quasiperiodic dynamics and, even, chaotic dynamics.

For the analysis of sequences of quantum averages extracted from the
iterations of the quantum neural map we calculate the following three
recurrence measures (Gonçalves, 2017, 2020):
\begin{itemize}
\item The recurrence probability: this is the number of diagonals below
the main diagonal with recurrence points, divided by the total number
of diagonals below the main diagonal in the recurrence matrix, since
the recurrence matrix is symmetric only the diagonals below the main
diagonal are counted, in this case, this metric provides for the probability
of finding a line with recurrence, in a random selection of diagonal
lines below the main diagonal.
\item The recurrence strength: this is the sum of the number of points that
fall within a distance no greater than the radius in each diagonal,
below the main diagonal, divided by the total number of diagonals
below the main diagonal with recurrence, this measure evaluates how
strong on average the recurrence is, if all lines with recurrence
had 100\% recurrence, for the radius chosen, then this number would
be equal to 1, the lower this statistic is, that is, the closer to
zero it is, the more interrupted the diagonals there are, which occurs
for stochastic systems and also for dererministic chaotic dynamics.
\item The conditional 100\% recurrence probability: this is the probability
that a diagonal line with recurrence has 100\% recurrence, for the
radius chosen.
\end{itemize}
These recurrence statistics can be used alongside the visual analysis
of a recurrence plot that plots the recurrence matrix (black and white
plot) or the distance matrix (colored recurrence plot), in the black
and white plot, which we will use in the present work, a point is
painted in black if it is a recurrence point and white if not, this
plot is a key element in addressing recurrence properties of both
low and high-dimensional dynamical systems (Gonçalves, 2017, 2020;
Eckmann \emph{et al.}, 1987; Gao and Cai, 2000), and have been employed
frequently in the analysis of neural network models as well as in
studies on brainwave dynamics (Thomasson \emph{et al.}, 2002; Acharya\emph{
et al.}, 2011; Aladag \emph{et al.}, 2010; Lopes \emph{et al.}, 2020),
in Gonçalves (2017, 2020) it was also applied to QRNN simulations.

In the present work, we will apply it to the study of the behavior
of the quantum neural activity field with a two-level neural firing
pattern, which can be built from the fermionic raising and lowering
operators on the Hilbert space $\mathbb{H}_{2}$, these raising and
lowering operators are defined as:
\begin{equation}
a=\left|0\right\rangle \left\langle 1\right|=\left(\begin{array}{cc}
0 & 1\\
0 & 0
\end{array}\right),\;a^{\dagger}=\left|1\right\rangle \left\langle 0\right|=\left(\begin{array}{cc}
0 & 0\\
1 & 0
\end{array}\right)
\end{equation}

For the anticommutator $\left\{ A,B\right\} =AB+BA$, these operators
obey the following relations:
\begin{equation}
\left\{ a,a^{\dagger}\right\} =1_{2}=\left(\begin{array}{cc}
1 & 0\\
0 & 1
\end{array}\right)
\end{equation}
\begin{equation}
\left\{ a,a\right\} =\left\{ a^{\dagger},a^{\dagger}\right\} =\left(\begin{array}{cc}
0 & 0\\
0 & 0
\end{array}\right)
\end{equation}
\begin{equation}
a^{\dagger}a\left|0\right\rangle =0\left|0\right\rangle 
\end{equation}

\begin{equation}
a^{\dagger}a\left|0\right\rangle =1\left|1\right\rangle 
\end{equation}

From the above equations, we can introduce the special case of a quantum
neural field operator which is the neural activity field operator
$N$ on the network defined as:

\begin{equation}
N(k)=1_{2}^{\otimes(k-1)}\otimes a^{\dagger}a\otimes1_{2}^{(n-k)},
\end{equation}
which has the eigenvalue spectrum obeying the following equation:
\begin{equation}
N(k)\left|...,s_{k},...\right\rangle =s_{k}\left|...,s_{k},...\right\rangle ,
\end{equation}
thus, considering the neural firing basis, at each neuron, the field
operator yields a value of 0 when the corresponding neuron is not
firing (not active) and of 1 when it is firing (active).

For the sequence $\left|\psi(t)\right\rangle $ with the expansion:
\begin{equation}
\left|\psi(t)\right\rangle =\sum_{s_{0},s_{1},...,s_{n-1}}\psi_{t}(s_{0},s_{1},...,s_{n-1})\left|s_{0},s_{1},...,s_{n-1}\right\rangle ,
\end{equation}
the quantum averages for the neural activity field at each neuron
coincide with the squared modulus of the quantum amplitudes, which
coincide, in turn, with the statistical measure for the neuron to
be active (firing): 
\begin{equation}
\begin{alignedat}{1}\left\langle N(k)\right\rangle _{t}=\left\langle \psi(t)\left|N(k)\right|\psi(t)\right\rangle =\\
=0\sum_{s_{0},...,s_{k-1},s_{k+1},...,s_{n-1}}\left|\psi(...,s_{k}=0,...)\right|^{2}+\\
+1\sum_{s_{0},...,s_{k-1},s_{k+1},...,s_{n-1}}\left|\psi(...,s_{k}=1,...)\right|^{2}\\
=\sum_{s_{0},...,s_{k-1},s_{k+1},...,s_{n-1}}\left|\psi(...,s_{k}=1,...)\right|^{2}
\end{alignedat}
\end{equation}

Using the $\mathbb{R}^{n}$ embedding $\left\langle N\right\rangle _{t}=\left(\left\langle N(0)\right\rangle _{t},\left\langle N(1)\right\rangle _{t},...,\left\langle N(n-1)\right\rangle _{t}\right)$,
$\left\langle N\right\rangle _{t}$ corresponds to the configuration
of the mean neural activity field at each neuron. As a final point
regarding the quantum neural activity field, it is relevant to stress
the relation between the neural activity field operator and the local
(neuron level) neural firing energy Hamiltonian operators, formally
these Hamiltonians can be defined as:
\begin{equation}
H_{k}=\omega\hbar N(k)
\end{equation}
where $\omega=2\pi f$ with $f$ corresponding to a neural firing
frequency in Hertz. Given the above equations, the energy eigenvalue
spectrum for the neuron is given by:
\begin{equation}
H_{k}\left|...,s_{k},...\right\rangle =\omega\hbar s_{k}\left|...,s_{k},...\right\rangle ,
\end{equation}
so that the energy is zero when the neuron is nonfiring and $\omega\hbar$
when the neuron is firing.

These operators commute and the total neural firing energy for the
network is given by:
\begin{equation}
H=\sum_{k=0}^{n-1}H_{k}=\sum_{k=0}^{n-1}\omega\hbar N(k)
\end{equation}
with the eigenvalue spectrum:
\begin{equation}
H\left|s_{0},s_{1},...,s_{n-1}\right\rangle =\omega\hbar\sum_{k=0}^{n-1}s_{k}\left|s_{0},s_{1},...,s_{n-1}\right\rangle 
\end{equation}

Having addressed the field theory, we now address the issue of the
local (neuron-level) von Neumann entropy dynamics.

\subsection{Entropy}

Formally, in a QuANN, due to the networked nature of the quantum computation,
the quantum dynamics at the local neuron level leads to entanglement
between the neurons' quantum dynamics, which means that, locally,
the neuron operates as an open quantum system. In general, with entanglement,
we cannot describe the neuron in terms of a vector, but rather by
a local density operator, tracing out the rest of the network's degrees
of freedom, which lead to the local (neuron-level) densities:

\begin{equation}
\rho_{k}(t)=Tr_{k}\left(\left|\psi(t)\right\rangle \left\langle \psi(t)\right|\right)
\end{equation}

The quantum information dynamics of the network, at the local neuron
level, can be addressed by calculating the von Neumann entropy with
a binary basis, indeed considering the general formula:
\begin{equation}
S(\rho)=-Tr\left(\rho\log_{2}\rho\right),
\end{equation}
which is equal to $0$ for a pure density, that is, a density given
by a projector $\left|\varphi\right\rangle \left\langle \varphi\right|$,
with $\left|\varphi\right\rangle $ being a normalized vector, we
can study the local neuron-level entropy dynamics employing similar
recurrence analysis techniques as those introduced in the previous
subsection, indeed, calculating the local von Neumann entropies for
each neuron in binary basis we get the entropy sequences:
\begin{equation}
S_{k}(t)=S(\rho_{k}(t))=-Tr\left(\rho_{k}(t)\log_{2}\rho_{k}(t)\right)
\end{equation}
that we can embed in $\mathbb{R}^{n}$, $\left(S_{0}(t),S_{1}(t),...,S_{n-1}(t)\right)$,
and to which we can apply the recurrence analysis techniques in order
to analyze the main dynamics for the local entropies.

When the input for the network is given by an initial pure density
(a projector) the entropy for the full network is zero and remains
zero under the evolution of the quantum neural map, since the map
is unitary and therefore does not change the global entropy, the local
neuron-level entropies, however, are not necessarily zero, due to
the entanglement dynamics associated with quantum networked computation.

Since the neuron-level densities are usually not equal to a projector,
there is ususally some level of entropy fluctuations at the local
neuron-level, indeed, the neuron-level networked dynamics tends to
a far-from-equilibrium dynamics that does not stabilize in a fixed
maximum entropy level, that is, while, in certain iterations, the
local neuron level's entropy can be led to close to the maximum entropy
this is not always the case, and there can be entropy reductions to
close to zero entropy, followed by entropy increases. In the case
of class 4 dynamics, for instance, we can also get class 4 dynamics
at the level of the entropy fluctuations themselves, with the entropy
fluctuating in a fluctuation band that is not maximal.

Having introduced the main concepts and framework, we now address
the example of the most elementary family of quantum recurrent neural
networks (QRNNs) the QRNNs comprised of two neurons characterized
by two-level neural firing activity (nonfiring and firing with a fixed
energy level).

\section{Complex Dynamics of a Quantum Recurrent Neural Network}

\subsection{Structure of the Network}

The most elementary QRNN is a network comprised of two neurons characterized
by a two-level neural firing activity, which, following the previous
section's formalism, is defined by:
\begin{equation}
G_{QRNN}=\left(\{n_{0},n_{1}\},\{(n_{0},n_{1}),(n_{1},n_{0})\},\mathbb{H}_{2}^{\otimes2},\Theta\left(\mathbb{H}_{2}^{\otimes2}\right)\right)
\end{equation}

The neural firing pattern basis for this network is given by:
\begin{equation}
B_{2}^{\otimes2}=\left\{ \left|0,0\right\rangle ,\left|0,1\right\rangle ,\left|1,0\right\rangle ,\left|1,1\right\rangle \right\} 
\end{equation}

The set of operators $\Theta\left(H_{Net}\right)=\left\{ U_{0},U_{1}\right\} $
is, in turn, defined such that each operator is a conditional unitary
operator that follows the neural connections, namely:

\begin{equation}
U_{0}=U_{0,0}\otimes\left|0\right\rangle \left\langle 0\right|+U_{0,1}\otimes\left|1\right\rangle \left\langle 1\right|
\end{equation}

\begin{equation}
U_{1}=\left|0\right\rangle \left\langle 0\right|\otimes U_{1,0}+\left|1\right\rangle \left\langle 1\right|\otimes U_{1,1},
\end{equation}
where $U_{r,s}$, for $r,s=0,1$, are elements of the unitary group
U(2).

From the above equations, it follows that, under the connection $(n_{0},n_{1})$,
the unitary gate $U_{1,1}$ is applied at the second neuron when the
first neuron is firing, while the unitary gate $U_{1,0}$ is applied
at the second neuron when the first neuron is nonfiring. In the reverse
direction, a similar conditional computation is performed, so that
when the second neuron is firing the computation at the first neuron
is given by the operator $U_{0,1}$ while, when the second neuron
is nonfiring, the computation at the first neuron is given by the
operator $U_{0,0}$.

Now, for a specific operator set $\Theta\left(\mathbb{H}_{Net}\right)$,
there are two possible activation orders for a unitary neural map
$U_{1}U_{0}$ or $U_{0}U_{1}$, the second alternative activates first
the connection $(n_{0},n_{1})$ and then the feedback (recurrent)
connection $(n_{1},n_{0})$, the first alternative activates first
the connection $(n_{1},n_{0})$ and then the feedback (recurrent)
connection $(n_{0},n_{1})$. In what follows, we will be working with
the activation sequence $U_{0}U_{1}$. 

The neural map that we will be analyzing has the following structure:
\begin{equation}
F=U_{0}U_{1}
\end{equation}
\begin{equation}
U_{1}=\left|0\right\rangle \left\langle 0\right|\otimes I+\left|1\right\rangle \left\langle 1\right|\otimes U_{r}
\end{equation}
\begin{equation}
U_{0}=I\otimes\left|0\right\rangle \left\langle 0\right|+U_{r}\otimes\left|1\right\rangle \left\langle 1\right|
\end{equation}

\begin{equation}
U_{r}=\cos\left(\frac{r\pi}{2}\right)\left(\left|0\right\rangle \left\langle 0\right|+\left|1\right\rangle \left\langle 1\right|\right)+\sin\left(\frac{r\pi}{2}\right)\left(-\left|0\right\rangle \left\langle 1\right|+\left|1\right\rangle \left\langle 0\right|\right)
\end{equation}
In matrix representation, the two operators $U_{1}$ and $U_{0}$
are given by:
\begin{equation}
U_{1}=\left(\begin{array}{cccc}
1 & 0 & 0 & 0\\
0 & 1 & 0 & 0\\
0 & 0 & \cos\left(\frac{r\pi}{2}\right) & -\sin\left(\frac{r\pi}{2}\right)\\
0 & 0 & \sin\left(\frac{r\pi}{2}\right) & \cos\left(\frac{r\pi}{2}\right)
\end{array}\right)
\end{equation}
\begin{equation}
U_{0}=\left(\begin{array}{cccc}
1 & 0 & 0 & 0\\
0 & \cos\left(\frac{r\pi}{2}\right) & 0 & -\sin\left(\frac{r\pi}{2}\right)\\
0 & 0 & 1 & 0\\
0 & \sin\left(\frac{r\pi}{2}\right) & 0 & \cos\left(\frac{r\pi}{2}\right)
\end{array}\right)
\end{equation}
which leads to the following matrix structure of the neural map $F$
\begin{equation}
F=\left(\begin{array}{cccc}
1 & 0 & 0 & 0\\
0 & \cos\left(\frac{r\pi}{2}\right) & -\sin^{2}\left(\frac{r\pi}{2}\right) & -\sin\left(\frac{r\pi}{2}\right)\cos\left(\frac{r\pi}{2}\right)\\
0 & 0 & \cos\left(\frac{r\pi}{2}\right) & -\sin\left(\frac{r\pi}{2}\right)\\
0 & \sin\left(\frac{r\pi}{2}\right) & \sin\left(\frac{r\pi}{2}\right)\cos\left(\frac{r\pi}{2}\right) & \cos^{2}\left(\frac{r\pi}{2}\right)
\end{array}\right)
\end{equation}

Considering the Schrödinger picture iteration: 
\begin{equation}
\left|\psi(t)\right\rangle =F\left|\psi(t-1)\right\rangle =U_{0}U_{1}\left|\psi(t-1)\right\rangle 
\end{equation}
and assuming the two expansions:
\begin{equation}
\left|\psi(t-1)\right\rangle =\sum_{s_{0},s_{1}}\psi_{t-1}(s_{0},s_{1})\left|s_{0},s_{1}\right\rangle 
\end{equation}
\begin{equation}
\left|\psi(t)\right\rangle =\sum_{s_{0},s_{1}}\psi_{t}(s_{0},s_{1})\left|s_{0},s_{1}\right\rangle 
\end{equation}
the amplitudes at $t-1$ and at $t$ are linked by:
\begin{equation}
\psi_{t}(0,0)=\psi_{t-1}(0,0)
\end{equation}
\begin{equation}
\begin{aligned}\psi_{t}(0,1)=\cos\left(\frac{r\pi}{2}\right)\psi_{t-1}(0,1)-\\
-\sin^{2}\left(\frac{r\pi}{2}\right)\psi_{t-1}(1,0)-\sin\left(\frac{r\pi}{2}\right)\cos\left(\frac{r\pi}{2}\right)\psi_{t-1}(1,1)
\end{aligned}
\end{equation}
\begin{equation}
\psi_{t}(1,0)=\cos\left(\frac{r\pi}{2}\right)\psi_{t-1}(1,0)-\sin\left(\frac{r\pi}{2}\right)\psi_{t-1}(1,1)
\end{equation}
\begin{equation}
\begin{aligned}\psi_{t}(1,1)=\sin\left(\frac{r\pi}{2}\right)\psi_{t-1}(0,1)+\\
+\cos\left(\frac{r\pi}{2}\right)\sin\left(\frac{r\pi}{2}\right)\psi_{t-1}(1,0)+\cos^{2}\left(\frac{r\pi}{2}\right)\psi_{t-1}(1,1)
\end{aligned}
\end{equation}

For $r=0$, $\psi_{t}(s_{0},s_{1})=\psi_{t-1}(s_{0},s_{1})$, so that
we get a class 1 dynamics, that is, a fixed point, since $\left|\psi(t)\right\rangle =\left|\psi(t-1)\right\rangle $,
so that the following holds:
\begin{equation}
\left|\psi(t)\right\rangle =\left|\psi(0)\right\rangle 
\end{equation}
for every normalized initial ket vector. With respect to the neural
activity field, introduced in the previous section, we get the quantum
averages:
\begin{equation}
\left\langle N(0)\right\rangle _{t}=\left|\psi_{0}(1,0)\right|^{2}+\left|\psi_{0}(1,1)\right|^{2}
\end{equation}
\begin{equation}
\left\langle N(1)\right\rangle _{t}=\left|\psi_{0}(0,1)\right|^{2}+\left|\psi_{0}(1,1)\right|^{2}
\end{equation}
which leads to a class 1 dynamics when the embedding in $\mathbb{R}^{2}$
is performed, since the dynamical point for the mean field is a fixed
point:
\begin{equation}
\left\langle N\right\rangle _{t}=\left(\left|\psi_{0}(1,0)\right|^{2}+\left|\psi_{0}(1,1)\right|^{2},\left|\psi_{0}(0,1)\right|^{2}+\left|\psi_{0}(1,1)\right|^{2}\right)
\end{equation}

On the other hand, when $r=1$, the dynamics is class 2. The reason
for this requires a closer look at the iteration steps, in this case
from $\left|\psi(t-1)\right\rangle $ to $\left|\psi(t)\right\rangle $
we get the transition sequence:
\begin{equation}
\begin{aligned}\left|\psi(t)\right\rangle =\psi_{t-1}(0,0)\left|0,0\right\rangle -\psi_{t-1}(1,0)\left|0,1\right\rangle -\\
-\psi_{t-1}(1,1)\left|1,0\right\rangle +\psi_{t-1}(0,1)\left|1,1\right\rangle 
\end{aligned}
\end{equation}
\begin{equation}
\begin{aligned}\left|\psi(t+1)\right\rangle =\psi_{t-1}(0,0)\left|0,0\right\rangle +\psi_{t-1}(1,1)\left|0,1\right\rangle -\\
-\psi_{t-1}(0,1)\left|1,0\right\rangle -\psi_{t-1}(1,0)\left|1,1\right\rangle 
\end{aligned}
\end{equation}
\begin{equation}
\left|\psi(t+2)\right\rangle =\left|\psi(t-1)\right\rangle 
\end{equation}

Therefore, we have a 3-cycle, that is, after three iterations, the
ket vector returns to the configuration in which it was three iterations
before, which leads to the embedded sequence:

\begin{equation}
\begin{aligned}\left\langle N\right\rangle _{t}=\\
=\left(\left|-\psi_{t-1}(1,1)\right|^{2}+\left|\psi_{t-1}(0,1)\right|^{2},\left|-\psi_{t-1}(1,0)\right|^{2}+\left|\psi_{t-1}(0,1)\right|^{2}\right)
\end{aligned}
\end{equation}
\begin{equation}
\begin{aligned}\left\langle N\right\rangle _{t+1}=\\
=\left(\left|-\psi_{t-1}(0,1)\right|^{2}+\left|-\psi_{t-1}(1,0)\right|^{2},\left|\psi_{t-1}(1,1)\right|^{2}+\left|-\psi_{t-1}(1,0)\right|^{2}\right)
\end{aligned}
\end{equation}
\begin{equation}
\left\langle N\right\rangle _{t+2}=\left\langle N\right\rangle _{t-1}
\end{equation}

While only class 1 and 2 dynamics are possible for these two parameters,
more complex dynamics arise when $0<r<1$, for different initial conditions,
as we now show.

\subsection{Network Simulations}

In figure 1, we show the simulated sequences for the mean neural activity
field values at each neuron, $\left\langle N(0)\right\rangle _{t}$
and $\left\langle N(1)\right\rangle _{t}$, when $r=0.0005$, for
$\left|\psi(0)\right\rangle =\left|+\right\rangle \otimes\left|+\right\rangle $,
with $\left|+\right\rangle =\left(\left|0\right\rangle +\left|1\right\rangle \right)/\sqrt{2}$. 

\begin{figure}[H]
\begin{centering}
\includegraphics[scale=0.5]{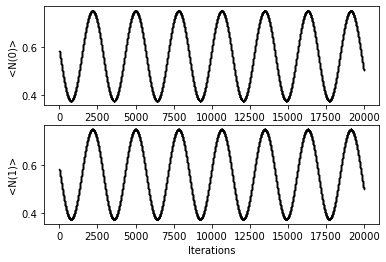}
\par\end{centering}
\caption{Simulation of the QRNN for $r=0.0005$, $\left|\psi(0)\right\rangle =\left|+\right\rangle \otimes\left|+\right\rangle $,
with 20,000 iterations after 10,000 initial iterations being dropped
for transients, the sample correlation between the two neurons' values
for the mean neural activity field is $0.99999977$.}
\end{figure}

In this case, we get an emergent pattern which follows a sinusoidal
curve for both neurons, with a high level of synchronization in the
sinusoidal pattern between the two neurons.

It is important to stress that the plot in figure 1 is a actually
a scatterplot, the appearance of a continuous periodic curve is due
to the close proximity of the dots, which means that the sinusoidal
pattern holds as an emergent pattern that appears in the dots' sequence
for each iteration. 

The sequence of dots is actually quasiperiodic, since for a recurrence
radius of 0 we get a recurrence probability equal to 0, however, the
quasiperiodicity is following an emergent continuous periodic curve,
that is, we get an emergent brainwave-like pattern that determines
the dynamics at the level of the mean field at each neuron which,
while being quasiperiodic, follows, in fact, an emergent continuous
periodic shape. Since the two emergent periodic curves are synchronized,
we get a pattern that is like an emergent synchronized neural wave
driving the network's dynamics.

While $r=0$ leads to a fixed point dynamics, increasing the parameter
to $r>0$, for low values of this parameter we get an emergent ``brainwave''
for different initial conditions.

In figure 2, we show the corresponding sequences of mean neural activity
field values at each neuron for the initial conditions $\left|\psi(0)\right\rangle =\left|0,1\right\rangle $,
$\left|\psi(0)\right\rangle =\left|1,0\right\rangle $ and $\left|\psi(0)\right\rangle =\left|1,1\right\rangle $,
and $r=0.0005$, in all three cases we get an emergent pattern that
is like a continuous wave, with a high correlation between the mean
field dynamics, however, for the first two cases, the sample correlation
is negative, which corresponds to a neural inhibitory dynamics, while
for the last case it is positive, corresponding to a neural reinforcing
dynamics. In the inhibitory cases this leads to an emergent negatively
correlated excitation-relaxation cycle between the two neurons with
a plateau in the excited phase and a smooth but faster relaxation
phase.

\begin{figure}[H]
\begin{centering}
\includegraphics[scale=0.4]{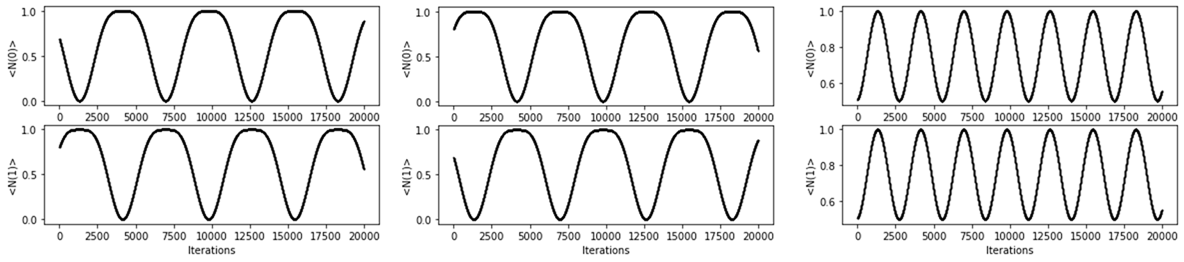}
\par\end{centering}
\caption{Simulations of the QRNN for $r=0.0005$, with 20,000 iterations after
10,000 initial iterations being dropped for transients, and with initial
conditions given by $\left|\psi(0)\right\rangle =\left|0,1\right\rangle $
(left), $\left|\psi(0)\right\rangle =\left|1,0\right\rangle $ (middle)
and $\left|\psi(0)\right\rangle =\left|1,1\right\rangle $ (right),
in the first case, the sample correlation is $-0.87907088$, in the
second $-0.87906459$ and in the third $0.99999955$}
\end{figure}

The dynamics, shown in figures 1 and 2 are also characterized by emergent
continuous periodic dynamics driving the neurons' entanglement dynamics
with a signature at the von Neumann entropy levels, which is different
from Everett's quantum automaton dynamics where the entanglement leads
to a fixed branching entanglement pattern (Everett, 1957, 1973), the
difference is that Everett's quantum automata theory is linked to
a formalization of a laboratory-based theory of quantum measurement,
and the measurement is a feedforward single interaction between the
observer and the observed system plus apparatus, in this sense at
the level of the observer's description there is an entanglement-related
local decoherence (Tegmark, 2000; Joos \emph{et al.}, 2003).

In QuANNs there is, however, no fixed/stable decoherence identifiable
as a von Neumann entropy rise to a fixed local maximally mixed density,
instead, due to the networked interaction, the entropy fluctuates,
so we do not have the same type of framework as is assumed in Everett
(1957, 1973) regarding quantum cognition, which, as stated, is an
expansion from a laboratory-based framework focused on a specific
type of interaction which is a quantum measurement, instead, for QuANNs,
we need to consider the issue of complex entropy dynamics associated
with interacting quantum systems in order to address quantum networked
processing at the local neuron level, where each neuron's computational
dynamics operates far-from-equilibrium.

In order to better understand this point, let us consider the local
neuron description. In the case of the above network, the local neuron-level
entropy, like the mean neural activity field, is also driven by an
emergent periodic continuous dynamics and exhibits a complex dynamical
relation with respect to the mean field values as we show in figure
3, for $\left|\psi(0)\right\rangle =\left|+\right\rangle \otimes\left|+\right\rangle $.
In this case, the initial entropy for each neuron was zero, since
the input vector for the network was separated into a tensor product
of two ket vectors. As the iterations proceed, we find that even though
the entropy sequence is discrete, like the mean neural activity field
values, it follows a continuous smooth curve which is periodic in
pattern, the entropy fluctuations range from near zero entropy which
corresponds to a pure density, to near 1, which corresponds to a maximum
entropy level associated with a depolarized mixed density, the mean
entropy being around 0.63 bits, as shown in table 1.

\begin{figure}[H]
\begin{centering}
\includegraphics[scale=0.5]{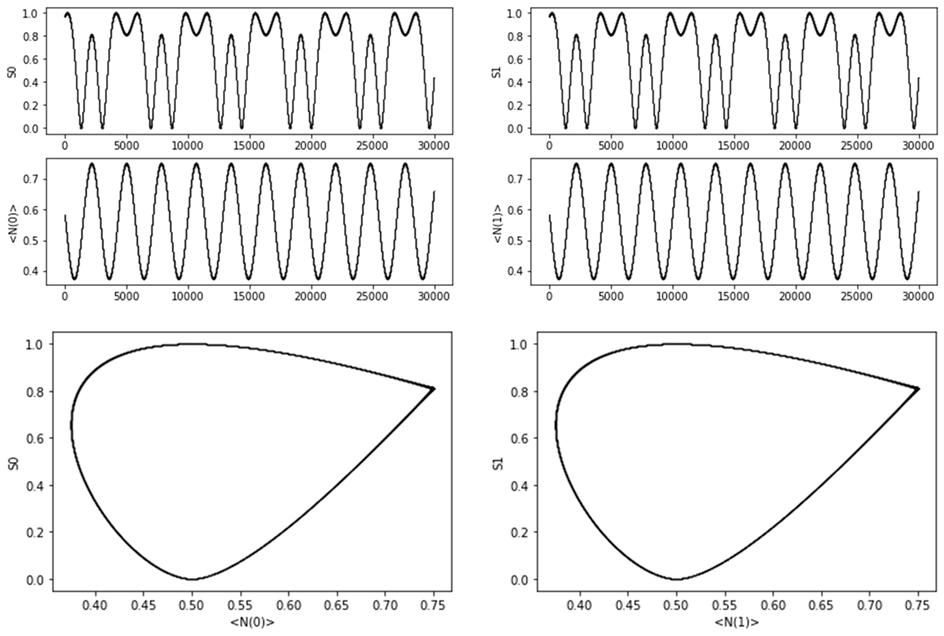}
\par\end{centering}
\caption{Simulations of the QRNN for $r=0.0005$, with comparison between the
mean neural activity field dynamics and the corresponding von Neumann
entropy values at neuron $n_{0}$ (left) and neuron $n_{1}$ (right),
with 30,000 iterations after 10,000 initial iterations being dropped
for transients, and with initial condition given by $\left|\psi(0)\right\rangle =\left|+\right\rangle \otimes\left|+\right\rangle $.}
\end{figure}

\begin{table}[H]
\begin{centering}
\begin{tabular}{|c|c|c|}
\hline 
 & $n_{0}$ & $n_{1}$\tabularnewline
\hline 
\hline 
Minimum Entropy & 8.0281238e-09 & 8.0278112e-09\tabularnewline
\hline 
Maximum Entropy & 0.9999998 & 0.9999998\tabularnewline
\hline 
Mean Entropy & 0.6347909 & 0.6347909\tabularnewline
\hline 
\end{tabular}
\par\end{centering}
\caption{Main entropy statistics for figure 3's simulation}

\end{table}

There is also a relation between the mean neural activity field at
each neuron and the respective entropy, in this case, we get an eye-like
structure, such that when the mean field value at the neuron is near
0.5, the entropy is either near 0 or near 1. For higher values of
the mean neural activity field, the dispersion in entropy fluctuations
diminish converging on a high but non-maximal entropy value for the
maximum mean neural activity field value, which also corresponds to
maximum mean energy at the neuron level.

Throughout the network's iterations each neuron is operating as an
open quantum computing system with fluctuations in entropy that can
range from a value close to zero to a value close to the maximum entropy
level, which illustrates the point that we do not get the basic fixed
decoherence pattern that is addressed in the context of quantum measurement
theory (Everett, 1957, 1973; Tegmark, 2000; Joos \emph{et al.}, 2003).
This is also the case for the entropy dynamics associated with the
initial conditions of figure 2.

Emergent smooth periodic curves driving the mean neural activity field
and entropy values are not the only patterns that are present in this
network's dynamics. When the network is initialized for $\left|\psi(0)\right\rangle =\left|+\right\rangle \otimes\left|+\right\rangle $,
as $r$ is increased, in the region of periodic emergent brainwave-like
patterns, the wavelength of the resulting neural waves decreases,
so that for $r$ very near 0, the wavelength is longer, but, as $r$
is increased, the wavelength decreases, as well as the sample correlation.
As shown in Gonçalves (2020), these correlations eventually transition
from from positive to negative values, with two main shapes characterizing
the dynamics of the mean neural activity field at neuron $n_{0}$
versus at neuron $n_{1}$, one with a triketa-like shape\footnote{It can be obtained from intersections of three ellipsoids.}
and the other with the shape of a trifolium, with the transition from
the triketa to the trifolium being progressive as $r$ is increased,
exhibiting complex quasiperiodic dynamics. Two examples of these attractors
are shown in figure 4.

\begin{figure}[H]
\begin{centering}
\includegraphics[scale=0.5]{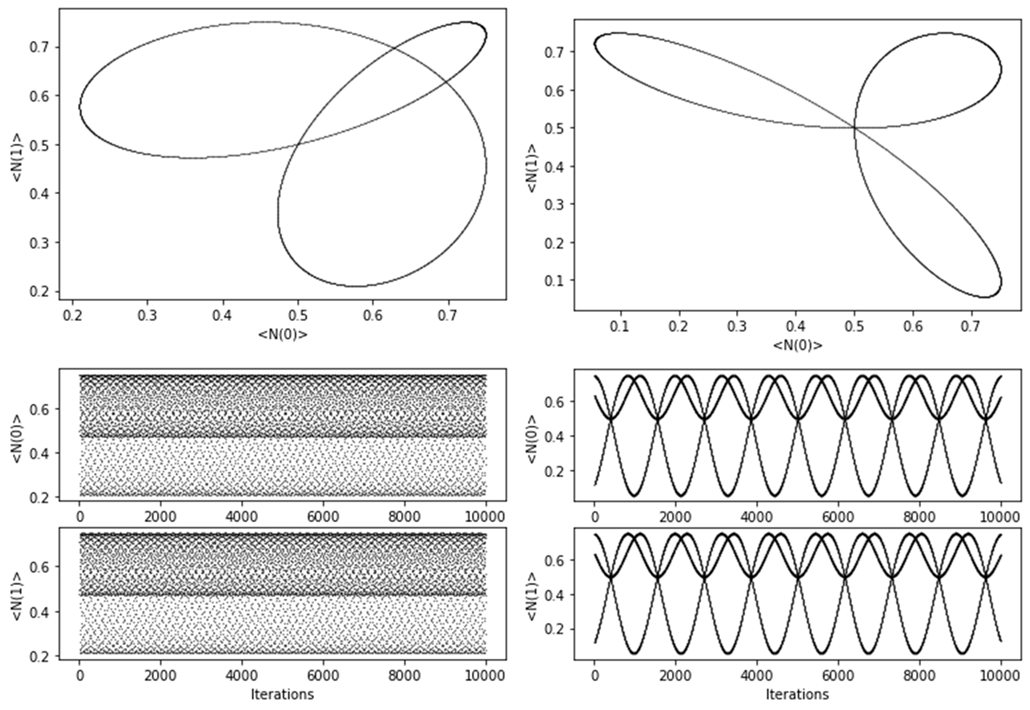}
\par\end{centering}
\caption{Attractor plots, showing the mean neural activity field at neuron
$n_{0}$ versus at neuron $n_{1}$, for $r=0.550129597$ (left) and
$r=0.999$ (right), with 10,000 iterations after 10,000 initial iterations
being dropped for transients and initial condition given, in both
cases, by $\left|\psi(0)\right\rangle =\left|+\right\rangle \otimes\left|+\right\rangle $,
also shown are the respective iterations graphs with the time series
sequences for the mean neural activity field values.}
\end{figure}

The attractor on the left corresponds to a value of $r$ where the
two neurons show a close to zero correlation\footnote{In this case, the sample correlation for figure 4's simulation is
1.85638176e-05.}, while, for the attractor on the right, the correlation is negative\footnote{In this case, the sample correlation for figure 4's simulation is
-0.49961075.}. While positive correlation indicates a dominance of an excitatory
dynamics between the two neurons, and a negative correlation indicates
the presence of an inhibitory relation, in the close to zero and negative
correlation region, unlike the positive correlation region, the dynamics
is characterized by complex quasiperiodic structures, mainly the triketa
which is the dominant geometrical structure, with the trifolium only
emerging for $r$ greater than 0.99. It is important to stress that
even though we get the same triketa shape, different values of the
parameter $r$ lead to different complex quasiperiodic patterns with
different recurrence structures as systematized in Gonçalves (2020).

The nonlinear relation between the sequences of mean field values
at each neuron emerges as $r$ increases, and the triketa becomes
the dominant emergent dynamical geometry both for near zero correlation
and negative correlation. In this sense, the linear correlation measure
becomes misleading as measure of the relation between the mean neural
activity field at each neuron.

This point becomes particularly relevant when considering the near
zero correlation case, since, while there is no dominant excitatory
or inhibitory dynamics, there is still a nonlinear relation between
the neurons which is characterized by a dynamics that exhibits, in
the recurrence structure, both signatures of dynamics with multiple
periodicities characterized by long resilient diagonals with 100\%
recurrence as well as broken diagonals and isolated dots that ususally
appear in stochastic or chaotic systems.

The dynamics is actually not chaotic nor periodic, instead, it is
closer, in regards to the recurrence structure, to the \emph{edge
of chaos}, such types of dynamics have been identified in other QuANN
models including recurrent networks interacting with an environment
(Gonçalves, 2017), as reviewed in the introduction. The recurrence
plots for the near zero correlation case are shown in figure 5, illustrating
this point.

\begin{figure}[H]
\centering{}\includegraphics[scale=0.65]{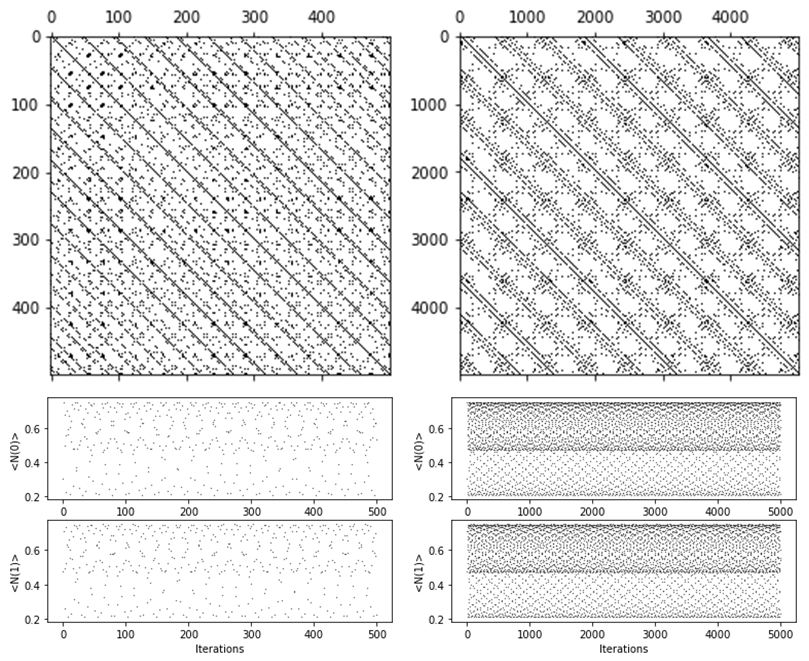}\caption{Sequence of mean field values at each neuron and recurrence plots
obtained for the ordered pairs $\left(\left\langle N(0)\right\rangle _{t},\left\langle N(1)\right\rangle _{t}\right)$,
$r=0.550129597$, 500 iterations (left) and 5,000 iterations (right),
after 10,000 initial iterations being dropped for transients, the
radius used for the recurrence plot was 0.1, the distance used was
the Euclidean distance, with recurrence points plotted in black and
initial condition given, in both cases, by $\left|\psi(0)\right\rangle =\left|+\right\rangle \otimes\left|+\right\rangle $.}
\end{figure}

As can be seen in figure 5, we no longer have the sequence of mean
field values following an emergent continuous periodic pattern, instead,
we get a dispersion in the form of a cloud of points, with the complex
pattern only being visible for higher number of iterations. The \emph{edge
of chaos} recurrence signatures show up upon an analysis of the recurrence
quantification measures. 

In table 2, we show the results from a recurrence analysis on the
ordered pairs $\left(\left\langle N(0)\right\rangle _{t},\left\langle N(1)\right\rangle _{t}\right)$,
for 20,000 iterations after 10,000 initial iterations being dropped
for transients, with $r=0.550129597$.

\begin{table}[H]
\begin{centering}
\begin{tabular}{|c|c|c|c|}
\hline 
{\small{}Radius} & {\small{}Recurrence Probability} & {\small{}Mean Recurrence Strength} & {\small{}P{[}100\% Rec|Rec{]}}\tabularnewline
\hline 
\hline 
{\small{}0} & {\small{}0} & {\small{}-} & {\small{}-}\tabularnewline
\hline 
{\small{}0.001} & {\small{}0.006750} & {\small{}0.116914} & {\small{}0.081481}\tabularnewline
\hline 
{\small{}0.01} & {\small{}0.069303} & {\small{}0.126819} & {\small{}0.086580}\tabularnewline
\hline 
{\small{}0.02} & {\small{}0.140057} & {\small{}0.129167} & {\small{}0.085684}\tabularnewline
\hline 
{\small{}0.03} & {\small{}0.211161} & {\small{}0.132081} & {\small{}0.085721}\tabularnewline
\hline 
{\small{}0.04} & {\small{}0.284464} & {\small{}0.134353} & {\small{}0.084725}\tabularnewline
\hline 
{\small{}0.05} & {\small{}0.360418} & {\small{}0.136247} & {\small{}0.083657}\tabularnewline
\hline 
{\small{}0.06} & {\small{}0.439772} & {\small{}0.137854} & {\small{}0.082661}\tabularnewline
\hline 
{\small{}0.07} & {\small{}0.525426} & {\small{}0.138656} & {\small{}0.080415}\tabularnewline
\hline 
{\small{}0.08} & {\small{}0.621031} & {\small{}0.138282} & {\small{}0.077939}\tabularnewline
\hline 
{\small{}0.09} & {\small{}0.737737} & {\small{}0.135577} & {\small{}0.073946}\tabularnewline
\hline 
{\small{}0.1} & {\small{}0.941097} & {\small{}0.123607} & {\small{}0.064502}\tabularnewline
\hline 
\end{tabular}
\par\end{centering}
\caption{Recurrence plot statistics for the ordered pairs $\left(\left\langle N(0)\right\rangle _{t},\left\langle N(1)\right\rangle _{t}\right)$,
obtained from 20,000 iterations simulations after 10,000 initial iterations
being dropped for transients, with $r=0.550129597$, initial condition
$\left|\psi(0)\right\rangle =\left|+\right\rangle \otimes\left|+\right\rangle $,
the statistics were calculated for increasing radius using a Euclidean
metric.}
\end{table}

The table shows how the recurrence statistics change with the increasing
radius, for radius 0, the recurrence probability is zero since there
are no diagonals, below the main diagonal, with recurrence, which
is indicative of an nonperiodic dynamics, the recurrence probability
and the mean recurrence strengths rise with the radius reaching a
94.1097\% recurrence probability for the radius 0.1, the probability
of finding a diagonal line with 100\% recurrence conditional on the
diagonal having recurrence points, however, rises initially with the
radius but then starts dropping, this indicates that the new recurrence
points that appear with the rise in radius are predominantly isolated
and clustered dots more characteristic of a noisy recurrence structure,
in this case, there is a resilient quasiperiodic skeleton of lines
with 100\% recurrence, but the remaining recurrence points do not
tend to produce a 100\% recurrence, which leads to a mix of an emergent
stochastic-like recurrence pattern intermixed with a few long diagonals,
characterizing a complex quasiperiodic dynamics, this is characteristic
of \emph{edge of chaos} signatures.

To evaluate the periodicities involved, we can calculate the distances
between the 100\% recurrence lines, these distances provide for an
evaluation of the quasiperiodic skeleton, in this case, for a radius
of 0.1, we find the present of three cycles, a 5 iterations cycle,
a 21 iterations cycle and a 26 iterations cycle, the fact that there
are different cycles present is characteristic of quasiperiodic dynamics,
in this case, there are two dominant cycles, for a radius of 0.1,
the first is the 21 iterations cycle, which occurs 836 times, followed
by the 5 iterations cycle which occurs 352 times, by contrast, the
26 iterations cycle only appears 25 times, as shown in table 3.

\begin{table}[H]
\begin{centering}
\begin{tabular}{|c|c|c|}
\hline 
Distances & Frequencies & \%\tabularnewline
\hline 
\hline 
5 & 352 & 29.019\%\tabularnewline
\hline 
21 & 836 & 68.920\%\tabularnewline
\hline 
26 & 25 & 2.061\%\tabularnewline
\hline 
\end{tabular}
\par\end{centering}
\caption{Distances between 100\% recurrence lines' statistics for the ordered
pairs $\left(\left\langle N(0)\right\rangle _{t},\left\langle N(1)\right\rangle _{t}\right)$,
obtained from 20,000 iterations simulations after 10,000 initial iterations
were dropped for transients, with $r=0.550129597$, initial condition
$\left|\psi(0)\right\rangle =\left|+\right\rangle \otimes\left|+\right\rangle $,
the statistics were calculated for a radius of 0.1 using a Euclidean
metric.}
\end{table}

Now, if we consider the von Neumann entropy dynamics, for $r=0.550129597$,
we need to consider the sequences of entropy values associated with
neurons $n_{0}$ and $n_{1}$, respectively, $S_{0}(t)$ and $S_{1}(t)$,
calculated from the respective neuron-level reduced densities, as
shown in figure 6 we also get a an nonperiodic dynamics, with multiple
diagonals but also a high number of interrupted diagonals and isolated
clusters, the power spectrum has multiple spikes at the high frequency
level for both neurons entropy sequences.

\begin{figure}[H]
\centering{}\includegraphics[scale=0.65]{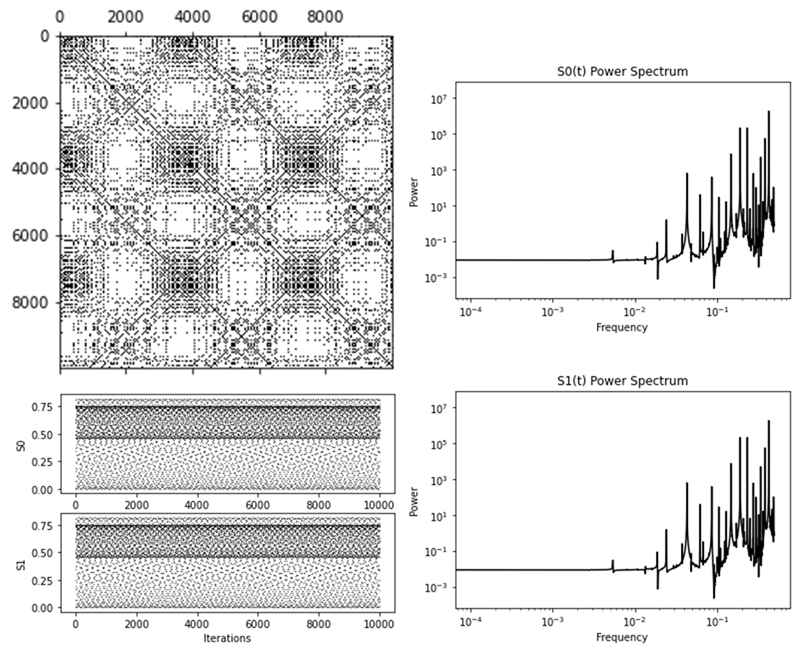}\caption{Sequences of entropy values $S_{0}(t)$ and $S_{1}(t)$ associated,
respectively, with neuron $n_{0}$ and $n_{1}$, with $r=0.550129597$,
recurrence plot obtained for the ordered pairs $\left(S_{0}(t),S_{1}(t)\right)$
and power spectrum associated with each entropy sequence, 10,000 iterations
were plottted, after 10,000 initial iterations being dropped for transients,
the radius used for the recurrence plot was 0.1, the distance used
was the Euclidean distance, with recurrence points plotted in black,
and initial condition given by $\left|\psi(0)\right\rangle =\left|+\right\rangle \otimes\left|+\right\rangle $. }
\end{figure}

There is, in this case, a predominant cycle which corresponds to a
47 iterations cycle that occurs 248 times with respect to the 100\%
recurrence lines at a radius of 0.1, the second cycle, in importance,
is a 68 iterations cycle, that occurs 88 times and a 115 iterations
cycle, that occurs 20 times, as shown in table 4.

\begin{table}[H]
\begin{centering}
\begin{tabular}{|c|c|c|}
\hline 
Distances & Frequencies & \%\tabularnewline
\hline 
\hline 
47 & 248 & 69.663\%\tabularnewline
\hline 
68 & 88 & 24.719\%\tabularnewline
\hline 
115 & 20 & 5.618\%\tabularnewline
\hline 
\end{tabular}
\par\end{centering}
\caption{Distances between 100\% recurrence lines' statistics for the ordered
pairs $\left(S_{0}(t),S_{1}(t)\right)$, obtained from 20,000 iterations
simulations after 10,000 initial iterations being dropped for transients,
with $r=0.550129597$, initial condition $\left|\psi(0)\right\rangle =\left|+\right\rangle \otimes\left|+\right\rangle $,
the statistics were calculated for a radius of 0.1 using a Euclidean
metric.}
\end{table}

Given the above results, we do not have, again, the stabilization
in a fixed entropy regime, the entropy fluctuations exhibit a complex
dynamics, associated with changes in entanglement levels and quantum
amplitudes. 

As shown in table 5, the lowest entropy value is close to zero, while
the highest entropy value is around 0.819 bits, no neuron is ever
led to the maximum entropy level associated with a depolarized mixed
density, and the mean entropy is around 0.498, so each neuron is operating
as a far-from-equilibrium open quantum system.

\begin{table}[H]
\begin{centering}
\begin{tabular}{|c|c|c|}
\hline 
 & $n_{0}$ & $n_{1}$\tabularnewline
\hline 
\hline 
Minimum Entropy & 2.2997412e-08 & 2.2997181e-08\tabularnewline
\hline 
Maximum Entropy & 0.8191482 & 0.8191481\tabularnewline
\hline 
Mean Entropy & 0.4976293 & 0.4976293\tabularnewline
\hline 
\end{tabular}
\par\end{centering}
\caption{Main entropy statistics for figure 6's simulation.}
\end{table}

For $r=0.999$, the entropy also has a complex pattern, as shown in
figure 7.

\begin{figure}[H]
\centering{}\includegraphics[scale=0.65]{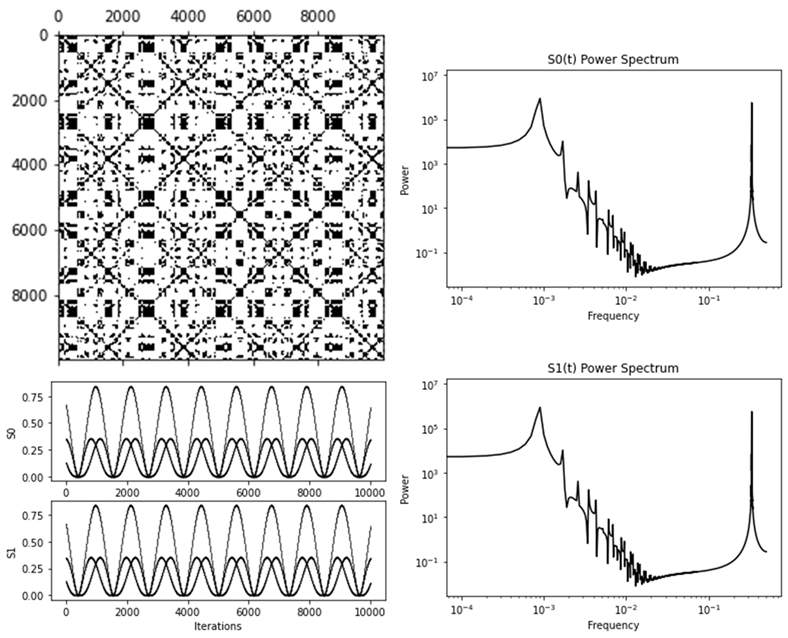}\caption{Sequences of entropy values $S_{0}(t)$ and $S_{1}(t)$ associated,
respectively, with neuron $n_{0}$ and $n_{1}$, with $r=0.999$,
recurrence plot obtained for the ordered pairs $\left(S_{0}(t),S_{1}(t)\right)$
and power spectrum associated with each entropy sequence, 10,000 iterations
were plottted, after 10,000 initial iterations being dropped for transients,
the radius used for the recurrence plot was 0.1, the distance used
was the Euclidean distance, with recurrence points plotted in black.}
\end{figure}

While, in figure 7, the entropy series values follow an emergent smooth
complex periodic curve, the process is very nontrivial, indeed, working
with the power spectrum, we find the presence of a power law decay
in the spectrum obtained for the von Neumann entropy sequences $S_{0}(t)$
and $S_{1}(t)$ with an additional rise for a significant peak at
the high frequency scale, indicating the presence of a strong periodicity
at that higher frequency. The recurrence plot shows elements of a
periodic skeleton but also local neighborhood fluctuations in the
sequence.

\begin{table}[H]
\begin{centering}
\begin{tabular}{|c|c|c|}
\hline 
 & $n_{0}$ & $n_{1}$\tabularnewline
\hline 
\hline 
Minimum Entropy & 9.5207835e-09 & 9.5208386e-09\tabularnewline
\hline 
Maximum Entropy & 0.8427277 & 0.8427277\tabularnewline
\hline 
Mean Entropy & 0.2608883 & 0.2608883\tabularnewline
\hline 
\end{tabular}
\par\end{centering}
\caption{Main entropy statistics for figure 7's simulation}
\end{table}

As shown in table 6, the interval of values for the entropy fluctuations
is similar to the previous one, the lowest entropy value is close
to zero, the highest entropy value is around 0.843 bits, the mean
entropy is, however, lower that in the previous case, around 0.261
bits, again, no neuron is ever led to the maximum entropy level and
each neuron is computing far from the maximum entropy level.

\section{Discussion}

In the present work we introduced a formalism for studying QuANNs
as complex quantum dynamical systems, demanding the introduction of
a quantum field theory on a quantum computing network and an expansion
of the concept of unitary map, worked within quantum chaos theory,
to the quantum computer science context of QuANNs and expanded further
to the quantum computing field theory.

The simulation of QuANNs as dynamical systems shows a diversity of
complex dynamics, even in small networks. For the most basic QRNN,
a network comprised of just two neurons, we obtained a diversity of
complex regimes in the quantum computational field dynamics that matches
the dynamical classes identified in classical networked computational
models studied within classical complexity sciences, and leads to
each neuron operating as a far-from-equilibrium open quantum system.

In the current work, we showed that dynamical classes, researched
upon in the classical complexity sciences in regards to networked
evolutionary computing systems' dynamics, characterize not only the
quantum mean neural activity field dynamics but also the local entropy
sequences, which differentiates between QuANNs operating as quantum
computing networked dynamical systems from another class of quantum
automata worked by Everett (1957; 1973) to address a type of measurement-like
interaction where the entropy for the local system rises to a fixed
level marked by a local diagonalization of the local density, which
has characterized the decoherence by interaction with the environment
literature (Tegmark, 2000; Joos \emph{et al.}, 2003). When linked
in network, each neuron operates as an open quantum computing system,
exhibiting entropy fluctuations that can get close to zero and, in
the case of the studied class 4 emergent quantum neural computing
field dynamics, never achieve a maximum entropy value.

Further research is needed into QuANNs as dynamical systems, both
in regards to the formalism of quantum computing field theory and
in regards to the simulation of these networks, especially with the
addition of more neurons and connections. From a computer science
standpoint, such a research may provide new results into low decoherence
far-from-equilibrium complex networked quantum computing systems with
possible applications in nanotechnology, quantum biology research,
quantum computing, quantum internet and A.I. research. Also, the far-from-equilibrium
class 4 dynamics leads to the emergence of a a form of resilient dynamical
memory encoded in the sequence of quantum averages, as resilient recurrences,
which may open up a research route into quantum dynamical memory storage.

Another implication of the results obtained from the simulations is
the need for a dialogue with neuroscience, considering especially
the fact of the emergence of brainwave-like patterns with different
wavelengths and the possibility of including network adaptive response
to signals leading to different wavelength responses at the quantum
neural activity level.

\section*{References}
\begin{description}
\item [{Acharya}] UR, Sree SV, Chattopadhyay S, Yu W, Ang PC. Application
of recurrence quantification analysis for the automated identification
of epileptic EEG signals. Int. J. Neural Syst., 2011, 21(3): 199-211.
\item [{Aladag}] CH, Egrioglu E, Kadilar C. Modeling Brain Wave Data by
Using Artificial Neural Networks. Hacettepe Jour. of Math. and Stat.,
2010, 39(1): 81-88. 
\item [{Beer}] K, Bondarenko D, Farrelly T, Osborne TJ, Salzmann R, Scheiermann
D, Wolf R. Training deep quantum neural networks. Nature Communications,
2020; 11: 808.
\item [{Braun,}] D. Dissipative Quantum Chaos and Decoherence. Springer,
Berlin, 2001.
\item [{Cramer}] JG. The Quantum Handshake: Entanglement, Nonlocality and
Transactions. Springer, Switzerland, 2016.
\item [{Dupuy}] J-P. The Mechanization of the Mind, Translation by MB DeBevoise,
Princeton University Press, Princeton, 2000.
\item [{Eckmann,}] J-P; Kamphorst, SO; Ruelle, D. Recurrence Plots of Dynamical
Systems. Europhys. Lett., 1987, 4(9): 973-977.
\item [{Everett}] H. 'Relative state' formulation of quantum mechanics.
Rev. of Mod. Physics, 1957; 29(3): 454-462. 
\item [{Everett}] H. The Theory of the Universal Wavefunction, PhD Manuscript,
In: DeWitt R and Graham N (eds.), The Many-Worlds Interpretation of
Quantum Mechanics. Princeton Series in Physics, Princeton University
Press, Princeton, 1973, 3-140.
\item [{Gao}] J and Cai H. On the Structures and Quantification of Recurrence
Plots. Phys. Lett., 2000, A270: 75-87.
\item [{Gonçalves}] CP. Quantum Cybernetics and Complex Quantum Systems
Science - A Quantum Connectionist Exploration. NeuroQuantology 2015a;
13(1): 35-48.
\item [{Gonçalves}] CP. Financial Market Modeling with Quantum Neural Networks.
Review of Business and Economics Studies 2015b; 3(4):44-63.
\item [{Gonçalves}] CP. Quantum Neural Machine Learning: Backpropagation
and Dynamics. NeuroQuantology, 2017; 15(1): 22-41.
\item [{Gonçalves}] CP. Quantum Robotics, Neural Networks and The Quantum
Force Interpretation. NeuroQuantology, 2019a; 17(2): 33-55.
\item [{Gonçalves}] CP. Quantum Neural Machine Learning - Theory and Experiments.
Aceves-Fernandez, M.A. (Ed.). Machine Learning in Medicine and Biology.
IntechOpen, London, 2019b: 95-118.
\item [{Gonçalves}] CP. Quantum Stochastic Neural Maps and Quantum Neural
Networks. Neurobiology eJournal 4(3), Computational Biology eJournal
4(4) and Information Systems eJournal 3(9) (SSRN), 2020; doi: 10.2139/ssrn.3502121.
Accessed date: February 2, 2022.
\item [{Gorodkin}] J, Sørensen A, Winther O. Neural Networks and Cellular
Automata Complexity. Complex Systems, 1993; 7:1-23.
\item [{Houssein}] EH, Abohashima Z, Elhoseny M, Mohamed WM. Machine learning
in the quantum realm: The state-of-the-art, challenges, and future
vision. Expert Systems with Applications 2022; 194: 116512.
\item [{Ivancevic}] VG, Reid DJ, Pilling MJ. Mathematics of Autonomy: Mathematical
Methods for Cyber-Physical-Cognitive Systems. World Scientific, World
Scientific, Singapore, 2018.
\item [{Joos}] E, Zeh HD, Kiefer C, Giulini D, Kupsch J, Stamatescu I-O
(Eds.). Decoherence and the Appearance of a Classical World in Quantum
Theory. Springer, Germany, 2003.
\item [{Kauffman}] SA and Johnsen S. Coevolution to the Edge of Chaos:
Coupled Fitness Landscapes, Poised States, and Coevolutionary Avalanches.
J Theor Biol 1991; 149:467- 505.
\item [{Kauffman}] SA. The Origins of Order: Self-Organization and Selection
in Evolution. Oxford University Press, New York, 1993.
\item [{Kwak}] Y, Yun WJ, Jung S, Kim J. Quantum Neural Networks: Concepts,
Applications, and Challenges. IEEE, \emph{Twelfth International Conference
on Ubiquitous and Future Networks (ICUFN)}, 2021:413-416.
\item [{Langton}] C. Computation at the Edge of Chaos: Phase Transitions
and Emergent Computation. Physica D 1990; 42:12-37.
\item [{Lopes}] MA, Zhang J, Krzemi\'{n}ski D, Hamandi K, Chen Q, Livi
L, Masuda N. Recurrence quantification analysis of dynamic brain networks.
Eur J Neurosci. 2021, 53: 1040--1059.
\item [{Novikov}] DA. Cybernetics: From Past to Future. Springer, Switzerland,
2016.
\item [{Packard,}] NH. Adaptation toward the edge of chaos. University
of Illinois at Urbana-Champaign, Center for Complex Systems Research,
1988.
\item [{Parisi}] L, Neagu D, Ma R, Campean IF. Quantum ReLU activation
for Convolutional Neural Networks to improve diagnosis of Parkinson\textquoteright s
disease and COVID-19. Expert Systems with Applications, 2022; 187:
115892.
\item [{Stöckmann}] H-J. Quantum Chaos - an introduction. Cambridge University
Press, UK, 2000.
\item [{Tegmark}] M. Importance of quantum decoherence in brain processes.
Phys Rev E, 2000; 61(4):4194-4206.
\item [{Thomasson}] N, Webber CL Jr, Zbilut JP. Application of recurrence
quantification analysis to EEG signals. Int. J. Comp. Appl., 2002,
9: 1-6.
\item [{Wolfram}] S. A New Kind of Science. Wolfram Media, USA, 2002.
\end{description}

\end{document}